\documentclass[twocolumn,prd,amssymb,amsmath,preprintnumbers,floatfix,aps,nofootinbib]{revtex4-1}
\usepackage{bm,color}
\usepackage{slashed} 
\usepackage{graphicx}
\usepackage{soul} 
\usepackage{multirow}
\usepackage{comment}
\usepackage{epstopdf} 
\usepackage{mathtools}

\usepackage[colorlinks=true
,urlcolor=blue
,anchorcolor=blue
,citecolor=blue
,filecolor=blue
,linkcolor=blue
,menucolor=blue
,linktocpage=true
,pdfproducer=medialab
,pdfa=true
]{hyperref}

\newcommand{\beq}{\begin{equation}}
\newcommand{\eeq}{\end{equation}}
\newcommand{\bea}{\begin{eqnarray}}
\newcommand{\eea}{\end{eqnarray}}
\newcommand{\ba}{\begin{array}}
\newcommand{\ea}{\end{array}}

\def\m1{M_1}
\def\m2{M_2}
\def\m3{M_3}

\def\ch10{\tilde \chi^0_1}

\def\gev{\,{\rm GeV}}

\def\Im{\,{\rm Im}}
\def\Re{\,{\rm Re}}
\def\to{\rightarrow}

\newcommand{\lsim}{\mathrel{\mathop{\kern 0pt \rlap
  {\raise.2ex\hbox{$<$}}}
  \lower.9ex\hbox{\kern-.190em $\sim$}}}
\newcommand{\gsim}{\mathrel{\mathop{\kern 0pt \rlap
  {\raise.2ex\hbox{$>$}}}
  \lower.9ex\hbox{\kern-.190em $\sim$}}}
\begin{document}
\title{\boldmath \bf \Large 
Interference in the $gg\to h\to \gamma\gamma$ On-Shell Rate\\
and the Higgs Boson Total Width 
}
\author{\bf John Campbell$^a$, Marcela Carena$^{a,b,c}$, Roni Harnik$^a$ and Zhen Liu$^a$} 

\affiliation{$a$ Theoretical Physics Department, Fermilab, Batavia, IL 60510, USA \\ 
$b$ Enrico Fermi Institute, University of Chicago, Chicago, IL 60637\\
$c$ Kavli Institute for Cosmological Physics, University of Chicago, Chicago, IL 60637}

\date{\normalsize  \today}

\begin{abstract}

We consider interference between the Higgs signal and QCD background in $gg\to h\to
\gamma\gamma$ and its effect on the on-shell Higgs rate. The existence of sizable strong
phases leads to destructive interference of about 2\% of the on-shell cross section in the
Standard Model. This effect can be enhanced by beyond the standard model physics. In
particular, since it scales differently from the usual rates, the presence of
interference allows indirect limits to be placed on the Higgs width in a novel way,
using on-shell rate measurements. Our study motivates further QCD calculations to reduce
uncertainties. We discuss potential width-sensitive observables, both using total and
differential rates and find that the HL-LHC can potentially indirectly probe widths of order
tens of~MeV.

\end{abstract}

\preprint{
FERMILAB-PUB-17-131-T
}


\maketitle








\section{Introduction}
The recent discovery of a Standard Model (SM)-like Higgs boson at the LHC opens a new era of research in particle physics.
The Higgs boson 
is directly connected to  the origin of mass of fundamental particles and the electroweak scale. 
Therefore, precision tests of the properties of the Higgs boson 
  provide a unique  window  into these basic questions.
At the LHC, the  current sensitivity in the cleanest Higgs boson channels is already exceeding expectations, while projections for future sensitivity  after the high luminosity LHC  (HL-LHC) run  are  at  the few percent level~\cite{Dawson:2013bba,ATL-PHYS-PUB-2014-016}. 
 In anticipation of the coming  era of high precision Higgs physics, 
small effects in Higgs production and decay rates should be carefully  scrutinized.  More importantly,  higher order effects that were previously  neglected or only partially considered, 
as well as new observables, 
may shed light on beyond the standard model effects encoded in the Higgs partial and total decay widths.

In this paper, we explore the physics potential for constraining  the SM Higgs total decay width from the change in on-shell Higgs rates due to interference effects between the Higgs signal and the QCD background. This change in rates requires the existence of a so-called strong phase in the amplitudes, that can be present both  in the Higgs signal and in the continuum background, as is the case in the SM. We shall demonstrate that,
the different scaling behavior between   the strong phase induced interference  and the Breit-Wigner parts of the on-shell Higgs rate may allow the placement of bounds on, or even measurements of, the Higgs boson total width.
Both theoretical and experimental uncertainties are the leading  limiting factors in this program. On the other hand, without  the strong phase induced interference effects, fits to on-shell Higgs rates can only place bounds on the total width by making definite theoretical assumptions~\cite{Duhrssen:2004cv,LHCHiggsCrossSectionWorkingGroup:2012nn,Dobrescu:2012td}. 

In the following we shall  focus on the process $gg\to h\to \gamma\gamma$, which will be measured with very high precision at the LHC. The
interference effect affecting  the Higgs production rate through  this process was estimated more than a decade ago in Refs.~\cite{Dicus:1987fk,Dixon:2003yb}. We explore this effect further and highlight for the first time  the resulting sensitivity to the Higgs boson total decay width.
We  study  the change in the cross section as a function of a veto on the  Higgs boson transverse momentum and as a function of the photon scattering angle in the diphoton rest frame.
In both cases we evaluate  the associated theoretical uncertainties, that certainly call for improvement.
More recently, interference effects in this channel have been studied extensively in~\cite{Martin:2012xc,Martin:2013ula,Dixon:2013haa,Coradeschi:2015tna}, putting  
 the main emphasis on the interference part proportional to the real component of the scalar propagator. This effect shifts the Higgs diphoton invariant mass peak at the $+10$~MeV to $-70$~MeV level, depending on the cuts.  Such a mass shift can provide sensitivity to the Higgs width, if the energy resolution is sufficiently good to allow for a comparably  accurate measurement of  the Higgs boson mass. The interference effect investigated in this work is complementary to those studies in the sense that, it is proportional to the imaginary component of the propagator and provides information on the Higgs total decay width from an on-shell Higgs boson cross section measurement. 
Our effect is also complementary to  the off-shell method~\cite{Kauer:2012hd,Caola:2013yja,Campbell:2013una}, since it is independent of new physics effects that may distort the measurements in the far 
off-shell region.

The rest of the paper is organized as follows. In Section~\ref{sec:result}  we discuss the basic nature of the strong phase induced interference effect between the  $gg\to h\to \gamma\gamma$ signal and the $gg\to \gamma\gamma$ SM background, and show how its dependence on the Higgs total width differs significantly from other Higgs rate observables. 
In Section~\ref{sec:NLO} we discuss results of our full next-to-leading-order (NLO) calculation, including the interference effects dominated by the background amplitude, considering both virtual and real emission contributions. 
We also present a discussion of novel observational aspects of this effect, including its angular and $p_T$-veto dependence, and outline strategies for constraining the Higgs boson total width in Section~\ref{sec:width}.
We summarize our results and possible future directions in Section~\ref{sec:conclude}.
In the Appendices, we provide details on the signal and background amplitude behaviors as well as on the input parameters used in our calculation. 

\section{interference effects and sensitivity to the Higgs Width}
\label{sec:result}

In order to quantify the effect of the Higgs boson on the diphoton production rate, including the effect of interference with QCD
background contributions, we compute observables based on the following combination of amplitudes,
\bea
|\mathcal{M}_h|^2 &=& |A_h+A_\mathrm{bkg}|^2-|A_\mathrm{bkg}|^2 \nonumber  \\ &=& |A_h|^2 + 2\,\mathrm{Re}\left[ A_h A^*_\mathrm{bkg}\right] \,,
\label{eq:sigrate}
\eea
where $A_h$ and $A_\mathrm{bkg}$ are amplitudes for diphoton production through an $s$-channel Higgs and for the rest of the SM processes, respectively.
For simplicity, helicity indices are suppressed in this section.
In the following, it is useful to write the amplitude for $gg \to h \to \gamma\gamma$ in a form which explicitly factors out the loop-induced couplings to gluons
($F_{gg}$) and photons ($F_{\gamma\gamma}$),
\begin{equation}
A_h \equiv A_{gg\to h\to\gamma\gamma}  \propto \frac {\hat s} {\hat s-m_h^2+i\Gamma_h m_h} F_{gg} F_{\gamma\gamma} \,.
\label{eq:Ah}
\end{equation}
Considering Eq.~(\ref{eq:sigrate}), the interference term can come from different contributions. 
Taking both $F_{gg}$ and $F_{\gamma\gamma}$ as real Wilson coefficients of effective vertices is sufficient for most purposes.
In such a case, $A_h$ is purely imaginary
when exactly on-shell, $\hat s=m_h^2$.  
In addition, the phase of the leading order
QCD background amplitude for $gg \to \gamma \gamma$ is often neglected.
These two approximations imply that the interference term in Eq.~(\ref{eq:sigrate}) vanishes
at $\hat s=m_h^2$.
Moreover, 
under the above conditions,
the interference term is proportional to the real part of $A_h$, and hence an odd function of $\hat s - m_h^2$. Therefore, upon $d\hat s$ integration the interference term does not change the overall rate, 
apart from a small contribution from the slope of the PDF.
The interference, however, does result in a small shift in the location of the peak in the $\gamma\gamma$ rate, that has been the subject of detailed
study~\cite{Martin:2012xc,Martin:2013ula,Dixon:2013haa,Coradeschi:2015tna}. 

Interestingly, a careful inspection of additional contributions to the interference term reveals effects that are not captured in the above discussion.
As explained in detail in appendices~\ref{sec:signalphase} and \ref{sec:bkgphase}, both the Higgs couplings $F_{gg}$ and $F_{\gamma\gamma}$ as well as the background amplitude $A_\mathrm{bkg}$
do receive absorptive contributions that arise from loops of particles that are sufficiently light to be on shell. 
The resulting phases that are induced are usually dubbed `strong phases' in the flavor literature and we will adopt this terminology here.~\footnote{Strong phases, which are CP even, get their name because they often arise in flavor physics from QCD dynamics. This is in contrast with CP odd weak phases which are Lagrangian parameters. A weak phase in Higgs physics, for example, would be the relative size of the Higgs couplings to $F\tilde F$ versus $FF$.}
In the presence of a strong phase we can write the interference term as
\begin{eqnarray}
|\mathcal{M}_h|^2_\mathrm{int} &&\equiv 2 \Re[A_h A_\mathrm{bkg}^*]= 
\frac {2 |A_\mathrm{bkg}||F_{gg}||F_{\gamma\gamma}|} {(\hat s - m_h^2)^2+\Gamma_h^2 m_h^2} \label{eq:phase}\\
&&\!\!
\times \!\left[ (\hat s-m_h^2) \cos (\delta_\mathrm{bkg}-\delta_h) \!+ \!m_h\Gamma_h \sin (\delta_\mathrm{bkg}-\delta_h) \right]\!\!, \nonumber
\end{eqnarray}
where we have taken $\delta_h=\mathrm{arg}[ F_{gg}]+\mathrm{arg}[F_{\gamma\gamma}]$ and $\delta_\mathrm{bkg}=\mathrm{arg}[ A_\mathrm{bkg}]$ as the signal and background strong phases, respectively.
The first term in the square bracket is the contribution to the interference term that, as we discussed below Eq.~(\ref{eq:Ah}), does not modify the overall rate once we integrate over $\hat s$. The second term is the subject of this work and leads to a modified rate in the presence of a signal or background strong phase. For convenience, we define $|\mathcal{M}_h|^2_\mathrm{int}=\mathcal{R}_h^\mathrm{int}+\mathcal{I}_h^\mathrm{int}$ and $\delta_s=\delta_\mathrm{bkg}-\delta_h$ such that
\bea
\mathcal{R}_h^\mathrm{int}&\equiv& \frac {2|A_\mathrm{bkg}||F_{gg}||F_{\gamma\gamma}|} {(\hat s - m_h^2)^2+\Gamma_h^2 m_h^2} (\hat s-m_h^2) \cos \delta_s \nonumber \\
\mathcal{I}_h^\mathrm{int}&\equiv& \frac {2|A_\mathrm{bkg}||F_{gg}||F_{\gamma\gamma}|} {(\hat s - m_h^2)^2+\Gamma_h^2 m_h^2} m_h\Gamma_h \sin \delta_s.
\eea

\begin{figure}[tbp]
\begin{center}
\includegraphics[scale=0.45,clip]{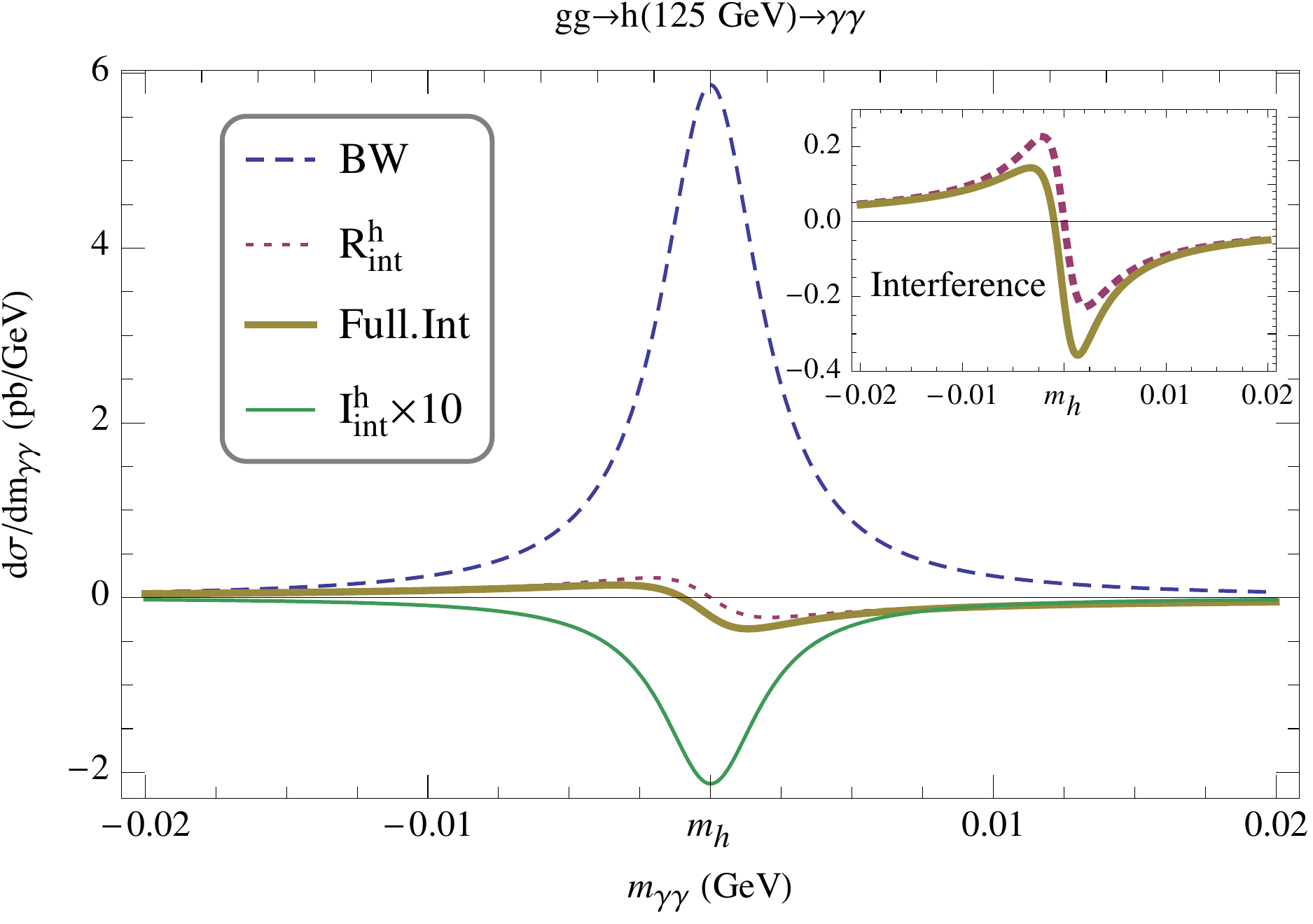}
\caption{The lineshape induced by various contributions to the cross-section for $gg \to h \to \gamma\gamma$ in the SM.
The Breit-Wigner line-shape, with no interference, is shown in blue (dashed) while the effect of $\mathcal{R}_h^\mathrm{int}$ and
$\mathcal{I}_h^\mathrm{int}$ (multiplied by a factor of 10) are shown in red (dotted) and green (solid), respectively.  The overall
effect of the interference in the full NLO calculation is given by the brown (solid) line. The insert in the top right is a magnification of the corresponding interference lineshapes.
}
\label{fig:sig_shape}
\end{center}
\end{figure}

In the SM the dominant contribution to $\mathcal{I}_h^\mathrm{int}$ comes from the phase of the background amplitude at two loops~\cite{Dicus:1987fk, Dixon:2003yb}. The signal amplitude also contains a
strong phase, mainly due to bottom quark loops. For a detailed discussion, we refer the reader to appendices~\ref{sec:signalphase}
and~\ref{sec:bkgphase}.  We have performed a calculation of the interference effect that accounts for both signal and background absorptive effects.
In Fig.~\ref{fig:sig_shape} we illustrate the features of the interference effects.
The line shape, the differential cross-section as a function of~$\hat s$, is shown
for the pure Breit-Wigner (only $|A_h|^2$), and for the interference contributions $\mathcal{I}_h^\mathrm{int}$ and $\mathcal{R}_h^\mathrm{int}$ as well as for the sum of both. 
For visualization, the interference contribution $\mathcal{I}_h^\mathrm{int}$ has been magnified by a factor of 10. 
In this figure we show the lineshapes obtained including NLO effects with virtual corrections only. After summing over different interfering helicity amplitudes, this amounts to averaged strong phases $\delta_h= (\pi+0.036)$ and $\delta_\mathrm{bkg}=-0.205$ for the signal and background, respectively.
In the next section we shall present a full NLO calculation and will find that the imaginary interference is destructive and
results in about a 2\% reduction in the overall rate.

Given that the interference term $\mathcal{I}_h^\mathrm{int}$ and the Breit-Wigner term have a different dependence on the Higgs boson total width, it follows that the on-shell cross section of $gg\to h\to \gamma\gamma$ gains sensitivity to the total width in a manner different than other on-shell Higgs cross sections that have negligible interference term contributions.
Schematically,
\bea
\sigma &\sim& \int d m_{\gamma\gamma} \,
\frac {|F_{gg}|^2|F_{\gamma\gamma}|^2} {(\hat s - m_h^2)^2+\Gamma_h^2 m_h^2}  \nonumber \\
& \times & \Bigl( 1 +
 \frac{2|A_\mathrm{bkg}|[(\hat s-m_h^2) \cos \delta_s + m_h\Gamma_h \sin \delta_s]}
   {|F_{gg}||F_{\gamma\gamma}|} \Bigr) \nonumber \\
 & \propto & \frac{|F_{gg}|^2|F_{\gamma\gamma}|^2}{\Gamma_h m_h}
  \left( 1 +  \frac{2 m_h\Gamma_h |A_\mathrm{bkg}| \sin \delta_s}
   {|F_{gg}||F_{\gamma\gamma}|} \right) 
   \,. \label{eq:sigma2}
\eea
This equation can be identified as expressing the parametric dependence of the cross section in the form of $\sigma=\sigma_\mathrm{BW}(1+\sigma_\mathrm{int}/\sigma_\mathrm{BW})$. 
The first term in this equation displays
the usual dependence of an on-shell cross section on the decay width, identical to that of the narrow-width-approximation result.  In the absence of the interference effect it
implies that such cross sections are insensitive to simultaneous changes to the Higgs couplings and total
width that leave the quantity $|F_{gg}|^2|F_{\gamma\gamma}|^2/\Gamma_h$ invariant -- this is the so-called flat direction in this
parameter space.  Combining the $\gamma\gamma$ rate with other channels will not eliminate this ambiguity since all on-shell rates exhibit an identical scaling with the corresponding couplings, namely $g_i^2 g_f^2 / \Gamma_h$ where $g_i$ and $g_f$ are the couplings for production and decay of the corresponding channel, respectively. However the presence of the second interference term in Eq.~(\ref{eq:sigma2}) lifts this degeneracy.
This special dependence on the total width can, in the fullness of time, be exploited by global fits of experimental data that
determine Higgs properties. 

\begin{figure}[tbp]
\begin{center}
\includegraphics[scale=0.45,clip]{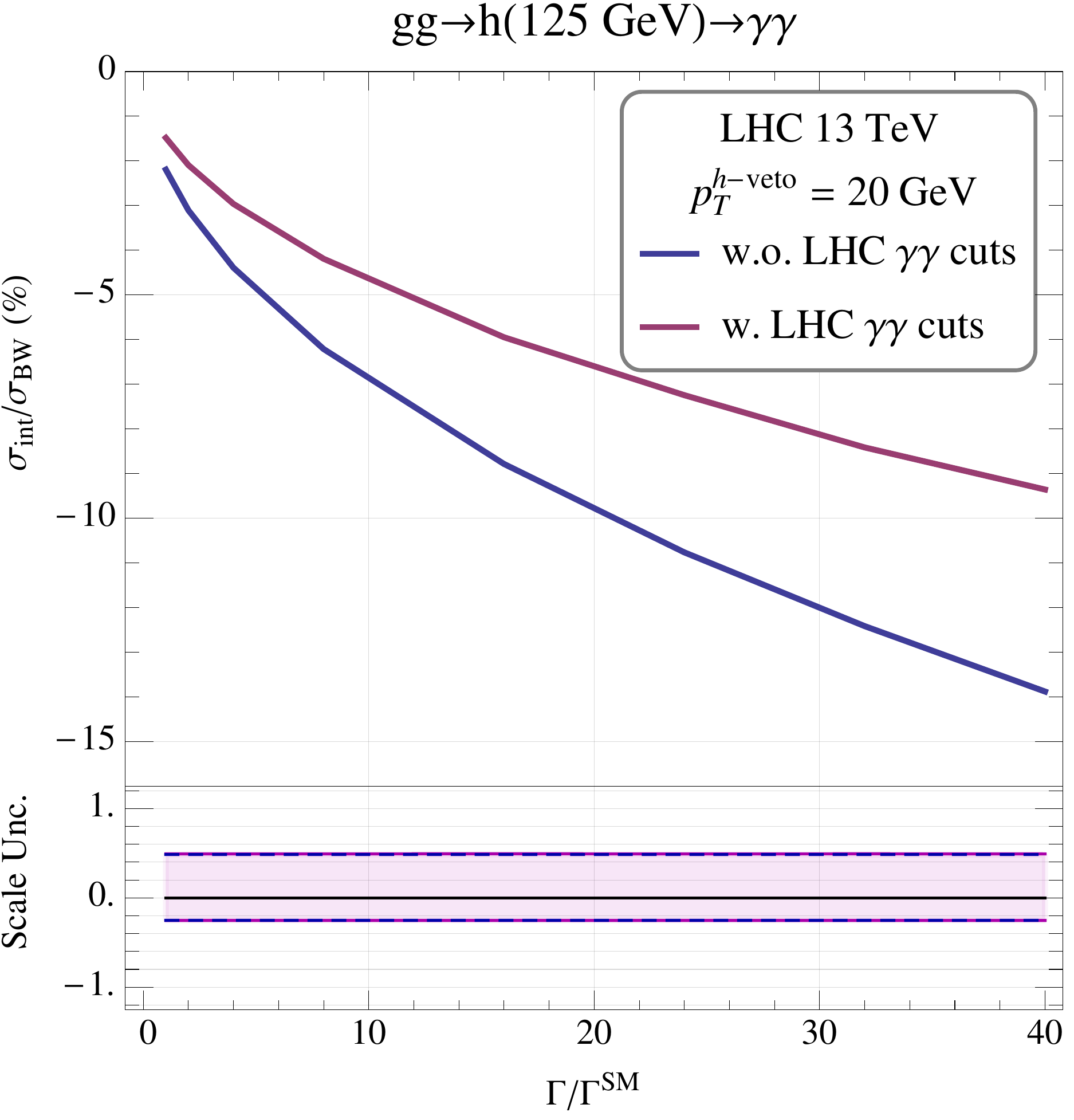}
\caption{
The total signal rate change due to the interference effect as a function of the Higgs total width normalized to its SM value, while keeping
the Breit-Wigner cross section identical to that of the SM Higgs. The magenta and blue (solid) lines represent the cases with and without LHC cuts
on the final state photons, respectively. The lower panel shows the scale variation uncertainties for these interference terms  as bands delimited by the blue (dashed)
 and magenta (solid) lines. The curves are obtained with a veto on the Higgs boson $p_T$  at 20 GeV. 
}
\label{fig:width}
\end{center}
\end{figure}

As a concrete example that demonstrates the potential of this novel effect, without loss of generality
we can consider excursions in the flat direction corresponding to,
\beq
\frac{|F_{gg}|^2|F_{\gamma\gamma}|^2}{|F^{\rm SM}_{gg}|^2|F^{\rm SM}_{\gamma\gamma}|^2}
 = \frac{\Gamma_h}{\Gamma^{\rm SM}_h} \,.
\label{eq:flatdir}
\eeq
The total Higgs cross section can then be written as,
\beq
\sigma = \sigma_\mathrm{BW}^\mathrm{SM} \left(1+ \frac {\sigma_\mathrm{int}^\mathrm{SM}} {\sigma_\mathrm{BW}^\mathrm{SM}}\sqrt{\frac {\Gamma_h} {\Gamma_h^\mathrm{SM}}}~\right)
\simeq \sigma_\mathrm{BW}^\mathrm{SM} \left(1-2\%\sqrt{\frac {\Gamma_h} {\Gamma_h^\mathrm{SM}}}~\right).\label{eq:deltasigma}
\eeq
%
The result of a full NLO calculation of the interference effect, the details of which will be presented in Section~\ref{sec:NLO}, are presented in Fig.~\ref{fig:width}, that shows the relative size of the interference effect as a function of the total width, normalized to its SM value, for parameter excursions defined by Eq.~(\ref{eq:flatdir}). The variation of the interference effect with the total width is shown imposing a 20~GeV $p_T^{h}$-veto, with and without LHC cuts on the final state photons (see Sec.~\ref{sec:NLO} for details). 
Since the interference
effect is largest at small scattering angles (see Fig.~\ref{fig:dsigmadtheta} in Sec.~\ref{sec:NLO}), the effect of the photon cuts is to reduce the expected interference.  This is only a small consideration in the SM, but leads to much
bigger differences for $\Gamma_h \gg \Gamma_h^{\rm SM}$.
Observe that in the SM the interference contribution is destructive. However,  if the sign of $F_{gg} F_{\gamma\gamma}$ were flipped, ($\delta_s\to \pi+\delta_s$), the interference effect would lead to an enhancement of the diphoton rate rather than a suppression. 
The theoretical scale uncertainty is shown in the bottom panel of Fig.~\ref{fig:width} and amounts to about 
$^{+50\%}_{-30\%}$,
as will be discussed in detail in the next section. For example, the interference effect is as large as $-(2.20^{+1.06}_{-0.55})\%$ without photon cut for SM Higgs.
Although a measurement at the 2\% level may be challenging at the LHC, 
this shows that a precise measurement of the $gg\to h \to \gamma\gamma$ rate can place a limit on the width of the Higgs boson. 
In this respect a measurement of the ratio of the $\gamma\gamma$ rate to the $4\ell$ rate is a promising route to reduce many of the systematic and theoretical, e.g. PDF and other parametric, uncertainties (see Sec.~\ref{sec:width} for details).


\section{Signal-Background Interference at NLO}
\label{sec:NLO}

The most recent calculation of the effect of the strong phase in the $gg\to h\to \gamma\gamma$  process was performed in
Ref.~\cite{Dixon:2003yb} in 2003. This calculation was primarily based on an analysis of the effect of the two-loop
contribution to the background amplitude, which is justified because of the size of the resulting strong phase at two-loop
relative to the one obtained in a one-loop analysis~\cite{Dicus:1987fk}.  
In this paper we perform a full NLO QCD
analysis of this effect, in which the singularities present in the two-loop amplitude are explicitly isolated
and cancelled by corresponding real radiation contributions.  We note that, as anticipated in the original analysis~\cite{Dixon:2003yb},
the impact of real radiation on the $\mathcal{I}_h^\mathrm{int}$ contribution to the interference is tiny.  However, the inclusion
of these effects allow us to present a consistent fixed-order calculation of the strong-phase induced interference, in a similar
fashion to a recent analysis of the effect of the $\mathcal{R}_h^\mathrm{int}$  contribution~\cite{Dixon:2013haa}.

For clarity we provide a description of the ingredients in our calculation.
At the LO level (one-loop in QCD) all quark mass effects are fully
taken into account. 
The NLO QCD corrections to the interference term
receive contributions from both the two-loop Higgs amplitude, which
is evaluated in the effective theory,
and the two-loop background amplitude which
is computed for five massless quark flavors circulating in the loop.  The corresponding
real radiation contribution is given by the interference of the one-loop diagrams
for the signal, $gg \to H(\to \gamma\gamma) g$ and background,
$gg \to \gamma\gamma g$~\cite{Bern:1993mq,Bern:2002jx} contributions.  Although they
are numerically negligible, for consistency and completeness we also include the
interference between the related $qg \to H(\to\gamma\gamma) q$ and $qg \to \gamma\gamma q$
(through a closed loop of quarks) processes.  As in the two-loop contributions, the
real radiation Higgs amplitudes are computed in the
effective theory and the background diagrams only include the effects of massless
quarks.  We do not include the effects of interference between one-loop $qg \to H(\to \gamma\gamma)q$ and tree-level
$qg \to \gamma\gamma q$ amplitudes since they do not contribute to the
strong phase induced interference
 term in the
effective theory~\cite{deFlorian:2013psa}. The singularities appearing in the calculation are handled using the
dipole subtraction method~\cite{Catani:1996vz} and the evaluation of all contributions is performed
using the parton-level code MCFM~\cite{Campbell:2011bn}. 
Our results have been obtained using the NLO CT14 PDF set~\cite{Dulat:2015mca},
with renormalization ($\mu_r$) and factorization ($\mu_f$) scales equal to $m_h$.  The
theoretical uncertainty on the predictions resulting from this scale choice is estimated
by performing a a six-point scale variation:
$\{\mu_r , \mu_f\} =\{ r m_h, f m_h \}$ with $r, f \in (\frac{1}{2}, 1, 2)$ and $rf \ne 1$.

Our NLO calculation allows the application of a realistic set of cuts that represent a typical LHC analysis
in this channel.   To do so, we impose a basic acceptance criterion for photons given by $p_T^{\gamma,\rm{hard}} > 40$~GeV, $p_T^{\gamma,\rm{soft}} > 30$~GeV and
$|\eta^{\gamma} | < 2.5$.  We also employ a simple isolation prescription, requiring that both photons are separated from a
parton with $p_T>3$~GeV by a distance $\Delta R > 0.4$.  The inclusion of real radiation contributions also allows a crude study of the effects of a jet (or, equivalently
at this order, Higgs) $p_T$ veto,
for which we reject events in which a parton is present with $p_T > p_T^{\rm veto}$ and $\eta < 3$.  We note that calculations of the effect of a jet veto on the pure Breit-Wigner
contribution are substantially more sophisticated and use perturbative input at N$^3$LO ameliorated with resumed predictions.  A summary of the current state-of-the-art
can be found in Ref.~\cite{deFlorian:2016spz}.  Nevertheless, the study performed here provides a consistent framework for studying the effect of the veto -- in which both
the interference and Breit-Wigner contributions are treated on the same footing -- which may be systematically improved in the future.


  
Our results for the full NLO calculation of the change in the rate relative to the Breit-Wigner Higgs cross section are
shown in Fig.~\ref{fig:width}.
The interference scales with the width
of the Higgs boson in accord with the expectation of Eq.~(\ref{eq:deltasigma}), up to small deviations induced by effects such as real radiation and PDFs.  
The theoretical uncertainty in this calculation shown in the bottom panel of Fig.~\ref{fig:width} is obtained following the scale variation procedure outlined above.
It is clear that this uncertainty is rather sizable and the dominant effect of the scale variation is captured by the change 
induced by the value of $\alpha_s(\mu_r)$.  
This is due to the fact that the strong phase is much bigger at two-loops than at one-loop, as discussed 
at length in Appendix~\ref{sec:bkgphase} and indicated in Fig.~\ref{fig:bkg_2g2a} therein.
This means that our NLO evaluation of this effect does not benefit from the substantial reduction in
scale dependence that might be expected from such a calculation;  instead, the prediction
for the interference has scale dependence similar to that of a LO calculation that is proportional to $\alpha_s^3(\mu_r)$.
Moreover,  this means that the prediction does not benefit from any reduction in scale dependence when expressed as
a fractional correction to the Higgs production cross section ($\sigma_{\rm int}/\sigma_{\rm BW}$).  This is due to the fact that
the scale dependence of $\sigma_{\rm BW}$ is already rather small, even in our NLO calculation.
A reduction of the uncertainty in $\sigma_{\rm int}$ would necessitate a three-loop calculation of a $2 \to 2$ scattering process, which is currently not tractable.  However, on the time-scale
over which the experimental precision could probe deviations at this level, i.e. the HL-LHC, there will surely be progress in this direction.




\subsection*{Kinematic Dependence of the Interference}
\label{sec:obs}

It is important to understand the variation of the interference effect with kinematic variables. 
%
%
\begin{figure}[tbp]
\begin{center}
\includegraphics[scale=0.44,clip]{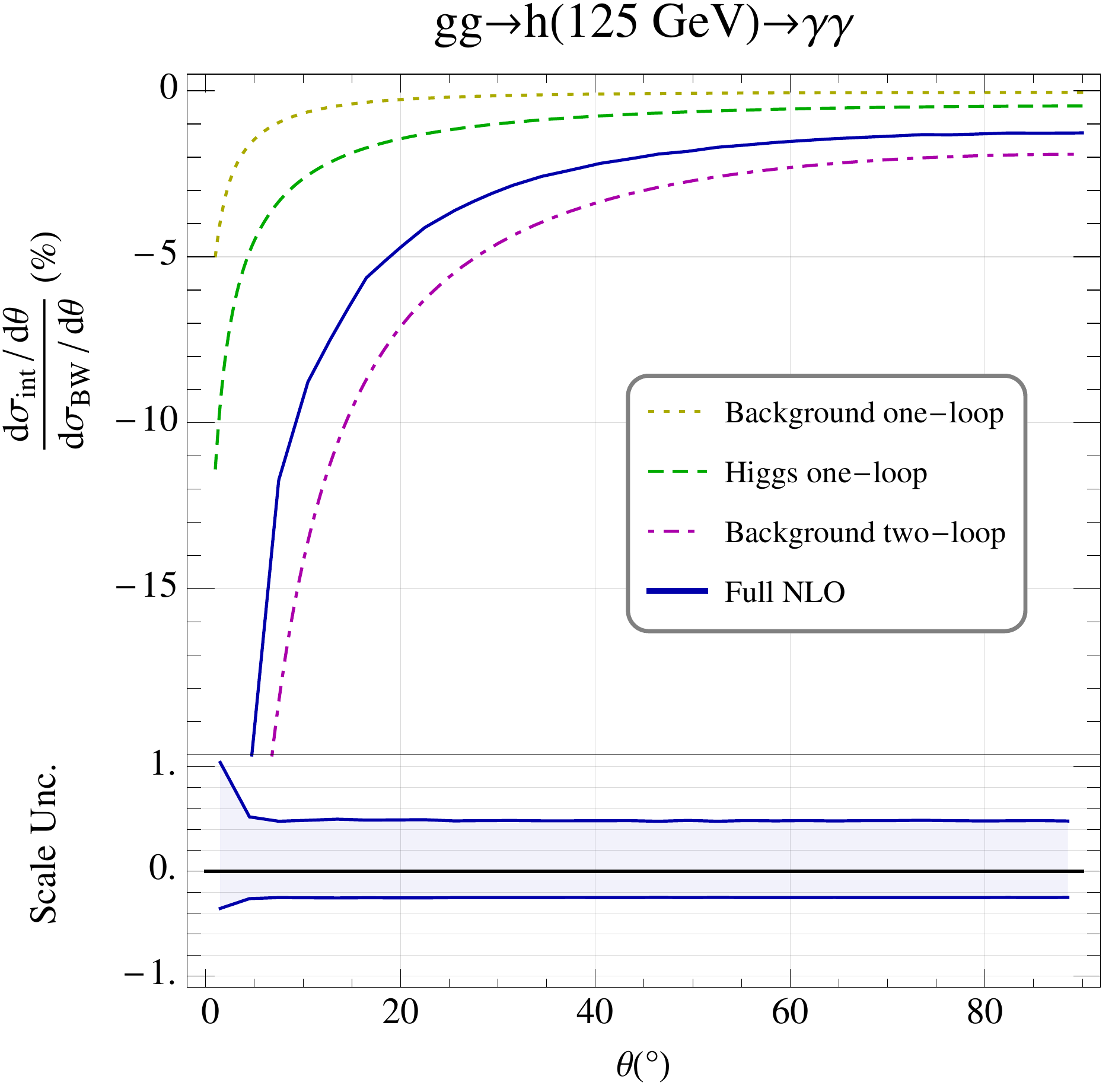}
\caption{Parton level cross section change due to the interference effect as a function of the photon scattering angle in the diphoton rest frame for the SM Higgs. The full NLO result is shown in solid blue curve. The blue band in the lower panel represents the scale uncertainty in the calculation of this effect. The dotted, dashed and dot-dashed lines correspond to partial calculations where the strong phase is included at various orders (see text). The partial  calculations include only virtual corrections while the full calculation result includes a Higgs $p_T$ veto of 20 GeV.}
\label{fig:dsigmadtheta}
\end{center}
\end{figure}
In Fig.~\ref{fig:dsigmadtheta} we show the differential distribution of the ratio $\sigma_\mathrm{int}/\sigma_\mathrm{BW}$ as a function of the photon scattering angle (in the $\gamma\gamma$ rest frame) for different orders of the signal and background amplitudes for the SM Higgs. The brown dotted line shows the interference effect at LO (1-loop) in both signal and background, but without including the signal strong phase. Staying at this order, but now including the strong phase in the Higgs amplitude leads to a somewhat larger effect, shown by the green dashed line. As discussed in Appendix~\ref{sec:bkgphase}, the background strong phase is suppressed by the masses of light quarks at one loop, but this suppression is absent at the two loop level.
We therefore see a sizable enhancement in this interference effect once we include the background amplitude at two loops (magenta, dot-dashed line). This curve is similar to the estimate of Ref.~\cite{Dixon:2003yb}. The full NLO calculation is shown by the solid blue line. The slight dilution of the effect going from the dot-dashed line to full NLO originates from an enhancement of $\sigma_\mathrm{BW}$ from real emission effects. For all curves we see that the interference effect is largest in the forward direction due to the kinematic behavior of the interfering background in this region, as shown in Fig.~\ref{fig:bkg_2g2a} in Appendix~\ref{sec:bkgphase}.

\begin{figure}[tbp]
\begin{center}
\includegraphics[scale=0.405,clip]{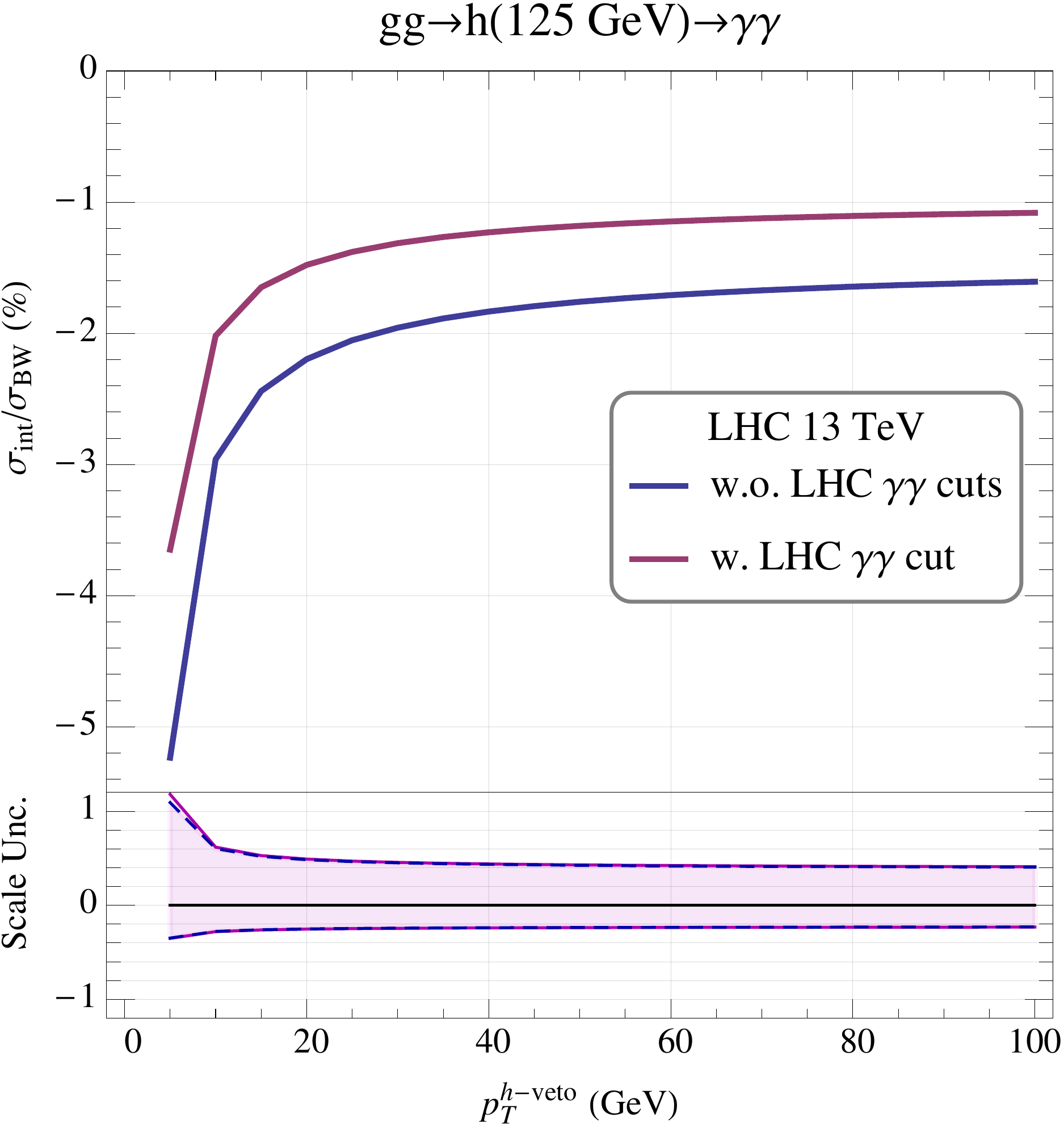}
\caption{
The cross section change due to interference effect in the SM, as a function of a veto on the Higgs boson $p_T$. The magenta and blue lines are for the cases  with and without LHC cuts on the final state photons, respectively. The  bands in the lower panel delimited by the magenta (solid)  and blue (dashed) lines represent the corresponding scale variation on these interference effect calculations.
}
\label{fig:ptveto}
\end{center}
\end{figure}

Fig.~\ref{fig:ptveto} shows the dependence of $\sigma_\mathrm{int}/\sigma_\mathrm{BW}$ as a function of a veto on the Higgs boson  $p_T$. 
Here we show the result both with and without LHC cuts on the final state photons described earlier in this section. 
As discussed earlier, real emission does not induce a sizable interference effect  given that it does not  yield additional sources of strong phase beyond those appearing in the LO amplitudes. 
In addition, its contribution enhances the Breit-Wigner cross section.
As a result, relaxing the Higgs $p_T$ veto leads to a larger contribution to $\sigma_\mathrm{BW}$ without contributing to $\sigma_\mathrm{int}$, diluting the $\sigma_\mathrm{int}/\sigma_\mathrm{BW}$ ratio.

\section{Strategies for probing the \\ Higgs  Boson Width}
\label{sec:width}

In the following we  comment on several important observational aspects and methods one can envision  to better measure the interference effect and constrain the Higgs boson width. 

The most straightforward approach is to compare the on-shell Higgs rate to the standard model prediction, with and without the inclusion of the interference term. As discussed in Sec.~\ref{sec:result}, the different parametric dependence of the Breit-Wigner and interference terms  would allows us to access the Higgs total width.
The HL-LHC projections for the statistical uncertainty in the rate of the 0-jet and 1-jet tagged Higgs to diphoton channels are 4\% and 5\%, respectively~\cite{ATL-PHYS-PUB-2014-016}. These measurements, however, are dominated by larger systematic and theoretical uncertainties, including those of beam luminosity and PDF. 
The best measured channels at the LHC, $gg\to h\to \gamma\gamma$ and $gg\to h \to 4\ell$, provide the most accurate cross section ratio, projected to 
 be measurable at the 4\% level~\cite{ATL-PHYS-PUB-2014-016}.   In contrast to single cross section measurements, the precision on this ratio is statistically limited. We observe, however, that if we naively scale the statistical uncertainty from the 8 TeV measurement~\cite{Khachatryan:2016vau} as the square root of the number of events, one reaches an impressive statistical uncertainty of 1-1.5\% at the HL-LHC. Together with the potential for improved analyses, it is not unreasonable to have high expectations for future HL-LHC performance.
Regardless of the ultimate reach of the HL-LHC, a few percent level is interestingly within striking distance of this novel interference effect. 

In light of the expectations for a remarkable experimental sensitivity, it is therefore highly motivated to improve the theoretical scale uncertainty. 
This, however, requires a full NNLO calculation of the interference effect, including a three-loop calculation of the $gg \to \gamma\gamma$ scattering amplitude.
Although this is a distant goal, among $2 \to 2$ calculations at this order, the amplitude for $gg \to \gamma\gamma$ would naturally be one of the first to be
computed due to its relationship to (a subleading-color part of) an all-gluon scattering amplitude.  Such calculations can, in general, be aided
by exploiting similarities between amplitudes computed in QCD and in super-Yang-Mills theories~\cite{Bern:1991aq}.
 
Keeping the current theoretical uncertainty band in mind, the projected sensitivity of 4\% on the ratio of $\gamma\gamma$ to $4\ell$ yields
can be translated into an upper  limit of 22, 14, and 8 on $\Gamma_h/\Gamma_h^\mathrm{SM}$ at 1-$\sigma$ level, for low, central and high theoretical expectations
on this interference effect, respectively. 
These expected sensitivities assume the observed measurement will agree with the standard model prediction, including our interference effect.
This also assumes that the couplings to photons and $Z$ bosons maintain their SM ratio and the photon and gluon couplings respect equation~(\ref{eq:flatdir}).
The Higgs cross section precisions are anticipated to improve by at least one order of magnitude at a future circular $pp$ collider~\cite{Arkani-Hamed:2015vfh,Contino:2016spe}. Using the same assumptions, this can be naively translated into  lower and upper limits on the Higgs total width of $0.5<\Gamma_h/\Gamma_h^\mathrm{SM}< 1.6$ at 1-$\sigma$ level using the central value from our NLO theory calculation. This high level of precision may thus establish the existence of the interference effect at 3-$\sigma$ level and differential distribution study will further improve it.
It is amusing to note that even today's $O(20\%)$ sensitivity to
Higgs rates, together with similar theoretical assumptions, places an indirect limit on the Higgs width of $\Gamma_h\lesssim 0.8~\gev$ at the 1-$\sigma$ level, which is competitive with the direct limit from the lineshape.

The interference effect on the total rate cannot be measured separately from the rate itself. This is why, in the preceding paragraph, we needed to make an assumption about the ratio of couplings. Otherwise, we would have to obtain such a ratio of couplings from other on-shell measurements or a global fit, adding additional uncertainties.
Therefore it is important to consider strategies for determining (or limiting) the size of the interference effect independent of  a flat direction or any additional theory input for the  ratio of the Higgs boson branching fractions.
The effects of interference can be measured independently from the rate by probing its dependence on kinematic observables such as the
scattering angle and the Higgs $p_T$(-veto) shown in Figs.~\ref{fig:dsigmadtheta} and \ref{fig:ptveto}, respectively. The best sensitivity
to this effect would presumably be achieved by employing a multi-variate analysis of the data. Here we will
simply consider a coarse binning of these distributions that can be compared with data such as Fig.~10(a) and Fig.~13(a)
of~Ref.~\cite{ATLAS-CONF-2016-067}, as the precision improves.

The isotropic nature of the Higgs boson decay means that, in principle, the interference effect can be mapped out by measuring the photon polar angle in the Higgs boson rest frame.  
Table~\ref{tab:differential} shows the size of the interference effect, as a fraction of the 
Breit-Wigner part of the Higgs boson signal, for a few bins in $\left|\cos\theta\right|$.  
Here we consider three selection criteria - the application of no cuts at all, no photon cuts but the
application of our Higgs $p_T$ veto, and finally our LHC photon cuts together with the veto.  
We will not attempt to estimate the reach of this analysis method.
However, observe that, for the $2 \to 2$ scattering configurations that dominate the interference effect,
the value of $\left|\cos\theta\right|$ is constrained by $\left|\cos\theta\right| < \sqrt{1-4(p_T^\gamma)^2/m_h^2} \approx 0.77$ for $p_T^\gamma > 40$~GeV. This explains 
the reduced effect of the interference in the 0.6-0.8 bin in the final column of Table~\ref{tab:differential} and the lack of an entry beyond that.
It would be possible to observe a significantly larger interference if the photon acceptance coverage were enlarged. 

\renewcommand{\arraystretch}{1.4}
\begin{table}[tbp]
\caption{The effect of the interference in bins of the photon
polar angle in the Higgs rest frame.  The effect is shown for three cases:  no cuts,
only a $p_T^h$ veto of 20~GeV, and LHC photon cuts in addition to the $p_T^h$ veto. The scale uncertainties are correlated across the various bins. Furthermore, the size of the effect grows uniformly in all table entries
as $\sim\sqrt{\Gamma/\Gamma_\mathrm{SM}}$, for parameter excursions in which all other
Higgs yields are held fixed. 
}
\begin{center}
\begin{tabular}{|c|c|c|c|}
\hline
 & \multicolumn{3}{c|}{$-{\sigma_{\rm int}}/{\sigma_{\rm BW}}$ (\%)} \\ \cline{2-4}
 $\left|\cos\theta\right|$ & ~~~ no cuts ~~ & ~~ $p_T^h$ veto ~~ & $\gamma\gamma$ cuts+veto \\ \hline
0.0--0.2
& $ 0.87_{-0.20}^{+0.34} $ 
& $ 1.28_{-0.32}^{+0.62} $ 
& $ 1.34_{-0.34}^{+0.68} $ 
 \\
0.2--0.4
& $ 0.91_{-0.21}^{+0.36} $ 
& $ 1.35_{-0.34}^{+0.65} $ 
& $ 1.41_{-0.36}^{+0.72} $ 
 \\
0.4--0.6
& $ 1.04_{-0.24}^{+0.41} $ 
& $ 1.53_{-0.38}^{+0.74} $ 
& $ 1.62_{-0.42}^{+0.83} $ 
 \\
0.6--0.8
& $ 1.37_{-0.31}^{+0.53} $ 
& $ 1.99_{-0.50}^{+0.96} $ 
& $ 1.65_{-0.40}^{+0.75} $ 
 \\
0.8--1.0
& $ 3.55_{-0.82}^{+1.45} $ 
& $ 4.85_{-1.23}^{+2.37} $ 
& -- 
 \\ \hline
0.0--1.0
& $ 1.52_{-0.35}^{+0.60} $ 
& $ 2.20_{-0.55}^{+1.06} $ 
& $ 1.48_{-0.38}^{+0.73} $ 
 \\
\hline
\end{tabular}
\end{center}
\label{tab:differential}
\end{table}%

One can also take the cross section ratio between vetoed and inclusive Higgs diphoton events as a probe of the interference effect since,
as shown in Fig.~\ref{fig:ptveto},
the size of the effect would be different in the two samples.  This would have the additional benefit of cancelling the leading theoretical and experimental uncertainties on the total cross sections.
We postpone any further phenomenological studies along these lines, which would require detailed collider simulation, to future work.

\section{conclusion}
\label{sec:conclude}

In this letter we discuss the change in the $gg\to h\to \gamma\gamma$ on-shell rate, due to interference between the Higgs signal and the QCD background amplitudes, as a way to provide a novel handle to constrain - or even measure~-  the Higgs boson total width. The key point here is  the different
scaling behavior, as a function of the Higgs boson width, between the interference  and the standard Breit-Wigner parts of the amplitude. The rate change is induced by the presence of strong phases, with the main effect coming from the background amplitude at two loops. We
perform a full NLO calculation at order $\alpha_s^3$ of the interference effect and find that in the standard model it leads to a reduction of the on-shell rate by $\sim 2\%$. In parameter excursions for which all other on shell rates are held fixed (to SM expectations or to the LHC measured best
fit), this effect grows with the square root of the Higgs total width. The scale uncertainty in our calculation is large, of order one of the original effect, calling for an improvement of the calculation of the QCD background beyond NLO. Such theoretical improvement, although challenging, is
justified given the prospects for a remarkable experimental sensitivity to some Higgs observables at the few percentage level.

We discussed the prospects for constraining the width by a  measurement of the ratio of the $gg\to h \to \gamma\gamma$ channel to that of the $gg\to h \to 4\ell$ channel. Although at present this ratio is statistically limited, future HL-LHC projections are quite promising, reinforcing the need for theoretical calculations of even higher precision.
We also explored kinematic observables such as the photon scattering angle in the diphoton rest frame and a veto on the  Higgs boson $p_T$,  which can be directly sensitive to the interference effect and provide complementary information to the total rate measurements. As an example, we show the variation of the interference effects as  a function of a coarse binning 
of the photon polar angle. More work is needed in this direction, that is beyond the scope of this paper, and multivariate analyses may prove to be the best approach to compare data with theoretical expectations in the future.

The proposed  method for gaining sensitivity to the Higgs boson width is complementary to other methods that have been discussed in the literature. As opposed to the off-shell method~\cite{Kauer:2012hd,Caola:2013yja,Campbell:2013una}, our effect is  active on-shell and is hence independent of  new physics effects that may distort the measurements in the far 
off-shell region. It is also qualitatively different from the mass shift method~\cite{Martin:2012xc,Martin:2013ula,Dixon:2013haa,Coradeschi:2015tna} since it directly affects the overall rate. 
It is interesting to imagine a scenario in which both on-shell effects are measured. In that case one can make a non-trivial test of the interference nature of these effects since the mass shift is proportional to  the real part of the background amplitude and the change in rate is proportional to the imaginary part. This information could in principle be compared to the $\gamma\gamma$ rate in the side-bands which is proportional to the magnitude of the background amplitude. Altogether our study aims at motivating  a more thorough examination of Higgs precision physics  taking into account the  strong phase induced interference effect in different Higgs boson observables.

\acknowledgments{Fermilab is operated by Fermi Research Alliance, LLC under Contract No. DE-AC02-07CH11359 with the U.S. Department of Energy.}

\appendix

\section{Signal amplitude and the strong phase}
\label{sec:signalphase}


The coupling of the Higgs boson to gluons can be parameterized as the coefficient $F_{gg}$
of the effective Lagrangian,
\beq
\mathcal{L}\supset \frac {\alpha_s} {4 \pi v} F_{gg} h G_{\mu\nu} G^{\mu\nu} \,,
\nonumber
\eeq
where $v$ is the SM Higgs vacuum expectation value,  taken as $v=246~\gev$, and the coupling is given by,
\beq
F_{gg}=\sum_q I_{1/2} (\tau_q)=\sum_q \frac 1 {\tau_q^2} \left[\tau_q + (\tau_q-1)f(\tau_q)\right]  \,.
\label{eq:fgg}
\eeq
Here $\tau_q={\hat s} /{4m_q^2} \approx {m_h^2} /{4 m_q^2}$ for a nearly on-shell Higgs boson and,
\beq
f(\tau) =\left\{
\begin{array}{lcl}
\arcsin^2(\sqrt\tau)&& {\rm for\ } \tau\leq 1,\nonumber\\ 
-\frac 1 4 \left(\log\frac {1+\sqrt {1-1/\tau}} {1-\sqrt {1-1/\tau}}-i \pi\right)^2\ &\ &{\rm for\ } \tau> 1
\end{array}\right.
\eeq

As expected from the optical theorem, the loop function quickly gains an imaginary component
after crossing threshold at $\tau_q=1$. This behavior implies that, although $F_{gg}$ is dominated by the top quark
loop that is purely real, all the light quarks ($b, c, s, d, u$) provide an imaginary component for the effective coupling $F_{gg}$.
The imaginary contribution from each light quark is linearly proportional to the Yukawa coupling, a suppression that can be seen by
examining the  $\tau_q \gg1$ limit of Eq.~(\ref{eq:fgg}) in which $F_{gg} \sim 1/\tau_q$.~\footnote{In the $\tau \gg 1$ limit, the imaginary part of the loop function $F_{gg}$ is proportional to $\pi/ \tau^2 (\tau
\log\tau)=(\log\tau)/\tau\rightarrow 1/\tau$ using L'H\^{o}pital's Rule. The $1/\tau$ is to be interpreted as $4 m_f^2/ \hat s = 2\sqrt{2} m_f y_f v
/m_f$ due to the parametrization of the loop functions, where one can see the Yukawa scaling explicitly.}
At the level of the cross-section the real components of these contributions  interfere destructively with the top quark contribution.

The three leading loop function contributions to the gluon-gluon-Higgs effective vertex are, after normalizing by the asymptotic value of $2/3$ at $\tau_q\to 0$,
\bea
F_{gg}^t&&=1.034, \nonumber\\
F_{gg}^b&&=-0.035+0.039 i, \\
F_{gg}^c&&=-0.004+0.002 i. \nonumber
\eea
It is useful to analyze a parameterization of the gluon-gluon-fusion production rate,
using the narrow-width approximation,
that exposes the dependence on the Yukawa couplings.  In terms of the ratios $\kappa_q \equiv y_q/y_q^{\rm SM}$,
in which the Yukawa couplings are expressed relative to their SM values,
\bea
\frac {\sigma_{\rm BW}} {\sigma_{\rm BW}^{\rm SM}}
= && 1.078 \kappa_t^2 -0.074 \kappa_t \kappa_b -0.008 \kappa_t \kappa_c  \nonumber \\
&&+ 0.003 \kappa_b^2+ O(<0.001; \kappa_t, \kappa_b, \kappa_c).
\nonumber
\eea
This equation reveals the destructive interference between the top quark loop contribution and the bottom and charm quark loops in the
effective gluon-gluon vertex.  Clearly the contribution from the square of the imaginary component is smaller than 1\% and
it is justifiable to ignore it when using the narrow-width-approximation.
The expression employed by the LHC ATLAS and CMS combination is~\cite{Khachatryan:2016vau},
\beq
\frac {\sigma_{\rm BW}} {\sigma_{\rm BW}^{\rm SM}}
= 1.06\kappa_t^2-0.07 \kappa_t \kappa_b + 0.01 \kappa_b^2.
\nonumber
\eeq
which agrees reasonably well with our expression using the input parameters specified in Eq.~(\ref{eq:input}).
We note here that the difference in the size of the bottom loop squared contribution ($\kappa_b^2$) is 
of the same order
as the charm loop-top loop interference ($\kappa_t \kappa_c$).
Moreover, if one takes into account such percent-level modifications then the interference effect discussed in this paper
-- at the level of around $-2$\% in the SM -- should also be taken into account. It is not captured in the present parametrization using the narrow-width-approximation.

The intrinsic strong phase for the gluon-gluon-Higgs interaction is not a small quantity, 
\beq
\arg(F_{gg})=0.042 \,.
\label{eq:argFgg}
\eeq
This means that the imaginary component of this effective vertex is as large as 4\% and can be important whenever there is a non-negligible
interference with a SM background. 

The coupling of the Higgs boson to photons is given at one-loop by,
\beq
F_{\gamma\gamma}=I_W(\tau_W)+\sum_{f} N_c^f e^2 Q_f^2 I_{1/2}(\tau_f) \,,
\eeq
where $I_{1/2}(\tau_f)$ is defined through Eq.~(\ref{eq:fgg}) and,
\beq
\tau_W= \frac {\hat s} {4m_W^2},\ I_W(\tau_W)=2+\frac 3 {\tau_W}+3 \left(\frac {2} {\tau_W}-1\right) f(\tau_W) \,.
\eeq
The contribution of the dominant $W$-boson loop flips the sign of the resulting strong phase and, in addition,
the effect is parametrically suppressed.  This results in $\arg(F_{\gamma\gamma})=\pi-0.006$.

Up to overall (irrelevant) phases, the one-loop helicity amplitudes for the process $gg\to h\to \gamma\gamma$ in the SM are,
\bea
A(++++)&&=A(----)=A(++--)=A(--++) \nonumber\\
&&=\frac {\alpha \alpha_s} { 8\pi^2 } \frac {m_h^2} {v^2} \frac {m_h^2} {s-m_h^2+i\Gamma_h m_h} F_{gg} F_{\gamma\gamma} \,.
\label{eq:sighel}
\eea
The amplitudes for all other helicity combinations are zero.
The overall strong phase for the signal amplitude apart from the Higgs propagator is then, 
\beq
\delta_h\equiv\arg (F_{gg} F_{\gamma\gamma}) =\pi+0.036.
\eeq
Comparing with Eq.~(\ref{eq:argFgg}) we can thus see that the strong phase from the gluon-gluon-Higgs vertex is reduced by an opposite-sign
strong phase from the photon-photon-Higgs vertex.

\section{The strong phase $\delta_{bkg}$ from the interfering one-loop and two-loop background}
\label{sec:bkgphase}

\begin{figure}[tbp]
\begin{center}
\includegraphics[scale=0.45,clip]{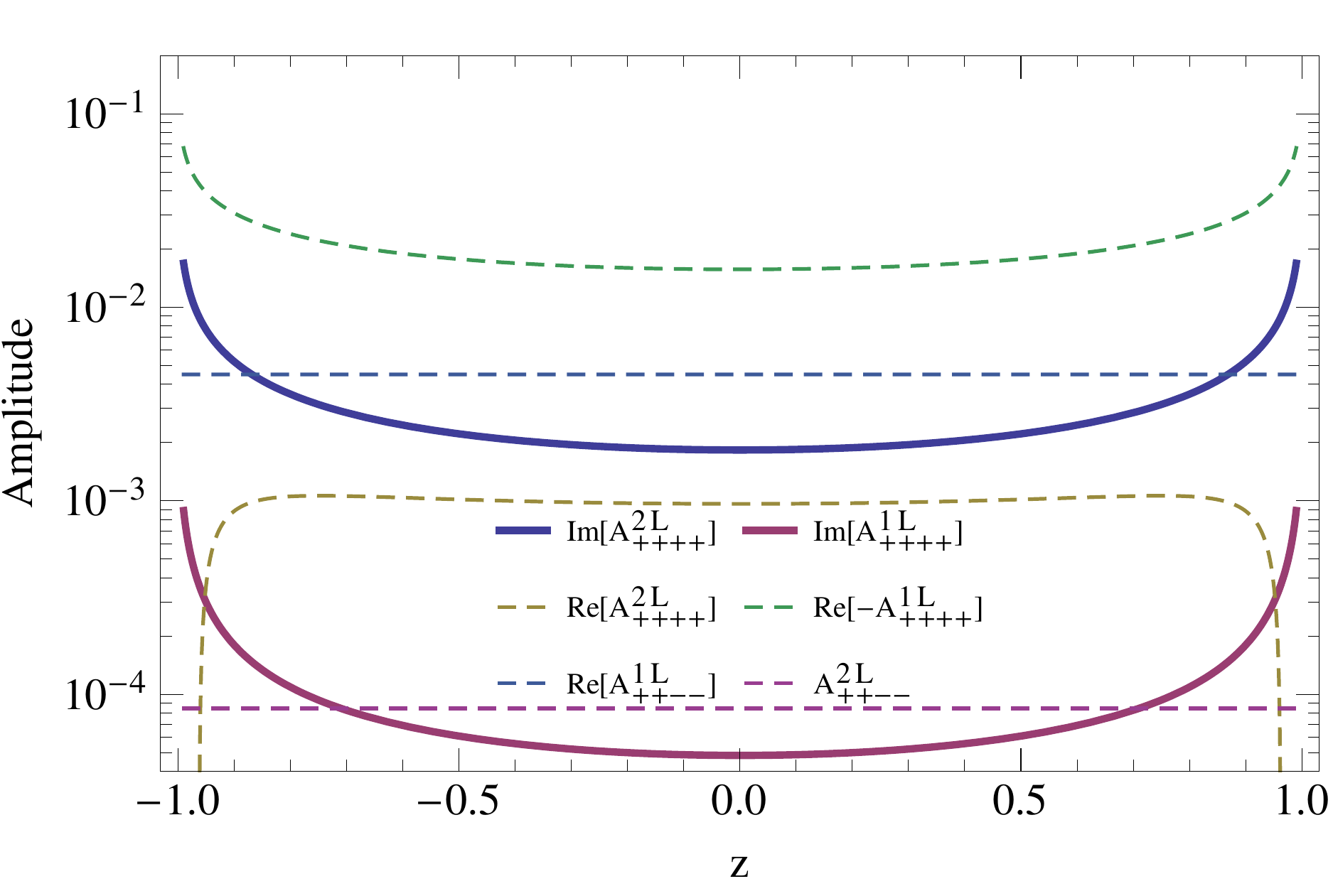}
\caption{Decomposition of the background amplitude of the SM process $gg\to \gamma\gamma$ at one-loop and two-loop order, at a center of mass energy $\sqrt{\hat {s}}=125~\gev$, as a function of the scattering angle, $z\equiv \cos\theta$, in the center of mass frame. The imaginary (real) components of the amplitudes are shown in solid (dashed) lines, respectively. }
\label{fig:bkg_2g2a}
\end{center}
\end{figure}


Since the amplitudes for the signal process $gg\to h\to \gamma\gamma$ are non-zero only for same-helicity gluons or photons,
as indicated in Eq.~(\ref{eq:sighel}), only four of the sixteen helicity amplitudes for the background amplitudes are relevant
when computing the interference contribution.  Further, parity can be used to relate two of these to their helicity-flipped
counterparts so that we need only discuss $A^{bkg}(++++)$ and $A^{bkg}(++--)$.

At both the one-~\cite{Bern:1991aq} and two-loop~\cite{Bern:2001df} levels, in the limit of massless quarks circulating
in the loop, the amplitudes $A^{bkg}(++--)$ are rational functions that, up to a phase, are simply constants.
Therefore they only receive absorptive contributions that are mass-suppressed~\cite{Bern:1995db}.  
The one-loop amplitude $A^{bkg,1L}(++++)$ only receives a mass-suppressed absorptive contribution, but it does depend
non-trivially on the kinematics of the process and diverges in the forward-scattering limit.  
The two-loop amplitude $A^{bkg,2L}(++++)$ does not have such a simple form and, even in the case of massless quarks, a non-trivial
absorptive contribution is present.  We note that other two-loop corrections, such as in the Higgs production and decay amplitudes,
do not induce any significant absorptive contributions.  Therefore it is the two-loop background amplitude that drives the strong
phase in this process. 

To illustrate these features, in Fig.~\ref{fig:bkg_2g2a}  we show the dispersive and absorptive 
components of the interfering background helicity amplitudes  at one-loop and two-loop order.   The amplitudes are shown as a function of the
cosine of the outgoing photon polar angle with respect to the beam direction, $z\equiv \cos\theta$, in the collision center of mass frame.
As anticipated, the magnitude of $A^{bkg}(++++)$ is bigger than that of
$A^{bkg}(++--)$ at both one- and two-loop order.  
The simple behavior of the interfering Higgs amplitudes allows us to gain a quantitative understanding of the overall strength of the interference effect
depicted in Fig.~\ref{fig:sig_shape}.   By summing over all interfering helicity amplitudes and averaging over $z$  we obtain the strong
phase of the background amplitudes, $\delta_{bkg}=-0.205$.

\section{Input parameters}

Since the strong phase from light fermions is proportional to the input mass parameter squared, their exact
values are crucial to computing the size of the expected effects.  
Both the running masses and couplings are evaluated at the Higgs mass, with values:
\bea
\overline m_t(m_h) &=& 169.76~\gev \nonumber\\
\overline m_b(m_h) &=& 2.9575~\gev \nonumber\\
\overline m_c(m_h) &=& 0.6169~\gev \nonumber\\
\alpha_S(m_h) &=& 0.11265 \nonumber\\
\alpha(m_h) &=& 1/127.5,
\label{eq:input}
\eea
The strong coupling is given by 2-loop running from the value determined in the PDF fit,
$\alpha_S(m_Z)=0.118$~\cite{Dulat:2015mca}.
We have explicitly checked that the contributions to the Higgs boson amplitude from light ($u$, $d$, $s$)
quarks are negligibly small, using values of 2.3, 4.6 and 95~MeV for up, down and strange quark masses, respectively.
We therefore set their masses to zero throughout.  The Higgs boson mass and total width are taken
as $125~\gev$ and $4.2$~MeV, respectively~\cite{Dittmaier:2011ti}.

\bibliographystyle{utphys}
\bibliography{refs_Higgs_phase}

\end{document}